\documentclass[referee]{aa}

\usepackage{graphicx}
\usepackage{txfonts}
\usepackage{color}

\usepackage{amstext}

\begin{document} 

\title{Collisional quenching of highly rotationally excited HF }

\author{Benhui Yang\inst{\ref{inst1}} \and K. M.~Walker\inst{\ref{inst1}}
    \and R. C.~Forrey\inst{\ref{inst2}}
    \and P. C.~Stancil\inst{\ref{inst1}} 
    \and N. Balakrishnan\inst{\ref{inst3}}}

\institute{Department of Physics and Astronomy and the Center for 
  Simulational Physics,  The University of Georgia, Athens, GA 30602, USA \label{inst1}
\and Department of Physics, Penn State University,
   Berks Campus, Reading, PA 19610, USA \label{inst2}
\and Department of Chemistry, University of Nevada, Las Vegas, NV 89154, USA\label{inst3}   }

\abstract
{Collisional excitation rate coefficients play an 
important role in the dynamics of energy transfer in the interstellar medium.
In particular, accurate rotational excitation rates are needed 
to interpret microwave and infrared observations of the interstellar gas 
for nonlocal thermodynamic equilibrium line formation.}
{Theoretical cross sections and rate coefficients for
collisional deexcitation of rotationally excited HF in the vibrational ground state are reported.}
{The quantum-mechanical close-coupling approach implemented in the nonreactive scattering code MOLSCAT was applied 
in the cross section and rate coefficient calculations on an accurate 2D HF-He potential energy surface.  
Estimates of rate coefficients for H and H$_2$ colliders were obtained from the HF-He collisional data
with a reduced-potential scaling approach.}
{The calculation of state-to-state rotational quenching cross sections for HF due to He with initial rotational 
levels up to $j=20$ were performed for kinetic energies from 10$^{-5}$ to 15000 cm$^{-1}$. 
State-to-state rate coefficients for temperatures between 0.1 and 3000 K are also presented. 
The comparison of the present results with previous work 
for lowly-excited rotational levels reveals significant differences. In estimating HF-H$_2$ rate coefficients, the
reduced-potential method is found to be more reliable than the standard reduced-mass approach.}
{ The current state-to-state rate coefficient calculations
 are the most comprehensive to date for HF-He collisions. 
We attribute the differences between previously reported and our results to differences in the adopted interaction potential energy surfaces. 
The new He rate coefficients can be used in a variety of applications. 
The estimated H$_2$ and H collision rates can also augment the smaller datasets previously 
developed for H$_2$ and electrons. 
}


\maketitle

\section{INTRODUCTION}

Molecular collisions, which are responsible for most of the excitation and reaction processes
involving molecules,
are important in the interstellar medium (ISM). Collisional excitation
and deexcitation processes compete with radiative transitions in 
populating molecular levels.  
In cold environments, the important collision partners are H$_2$ and He because of their large abundances,
except in photodissociation regions (PDRs) and diffuse gas where collisions
with electrons and H can become important.  
Precise laboratory data including collisional deexcitation rate coefficients
are required for a range of temperatures
to interpret the complicated interstellar spectra of molecular gas not in 
local thermodynamic equilibrium (LTE).
Because of the complexity and difficulty of direct measurements, only limited 
state-to-state collisional rate coefficients have been
measured for systems of astrophysical interest \citep[see, for example,][]{bru02}.
Therefore, astrophysical modeling heavily depends on theoretical prediction 
\citep[e.g.,][]{flo07,fau12,wie13,rou13,yang13,dub13}.

In this paper, we consider hydrogen fluoride (HF), an interstellar molecule containing 
a halogen element, which was first detected in the ISM by \citet{neu97}.
The reactivity of HF is weak, but it may be formed by the exoergic process 
F+H$_2$$\rightarrow$HF+H. 
Experimental rate coefficients for this
reaction were recently reported by \citet{tiz14} at temperatures between 10 and 100 K. 
 As a result of its structural stability and radiative properties, the HF molecule can be an alternative tracer to H$_2$ 
in diffuse regions;
HF is also the main reservoir of fluorine in the ISM \citep{mon11,van12a}.  
\citet{mon14} reported observations of HF in two luminous  galaxies, NGC 253 and NGC 4945,
 using the Heterodyne Instrument for the Far-Infrared (HIFI) on the {\it Herschel Space Observatory}.
With {\it Herschel}/HIFI, \citet{neu10} detected HF in
absorption from the ground rovibrational state. 
\citet{phi10}  presented a detection of the fundamental rotational 
transition of hydrogen fluoride in absorption toward Orion KL using {\it Herschel}/HIFI. 
The emission in the $j=1\rightarrow 0$ rotational transition of HF has been observed in the carbon
star envelope IRC +10216 by \citet{agu11}.
\citet{mon11} reported the first detection of HF toward a high-redshift quasar at $z=2.56$, while
\citet{van12b} presented observations of the HF $j = 1\rightarrow 0$ line in emission 
towards the Orion Bar.

The HF-He scattering system has been studied theoretically
and experimentally \citep{lov90,mos94,mos96,cha96,sto03,ree05,faj06}. 
The availability of ab initio HF-He potential energy surfaces (PESs) has stimulated  
theoretical studies of HF excitation due to He impact. 
 \citet{lov90} reported the first experimental study of the near-infrared 
vibrational-rotational spectra of the HeHF and HeDF complexes in a supersonic expansion.
The HeHF (HeDF) spectra showed good agreement with the prediction obtained using the
Hartree-Fock dispersion (HFD) type rigid-rotor potential of Rodwell et al. (1981).
The spectroscopic data were analyzed and used to probe the isotropic and anisotropic intermolecular
potentials of the complexes.
The two-dimensional (2D) interaction potential of \citet{mos94} was developed
from ab initio calculations using symmetry-adapted perturbation theory (SAPT).   
This SAPT potential is in good agreement with the empirical PES of \citet{lov90}. 
All transition frequencies obtained from the bound-state calculations using the SAPT potential
showed excellent agreement with the experimental spectra.
The SAPT potential
 has a global minimum for the linear He-HF arrangement and a secondary minimum
for the linear He-FH geometry.
The accuracy of the SAPT potential was also confirmed by agreement between 
calculated differential and integral cross sections on a slightly modified
 SAPT potential and experimental results \citep{mos96}. 
 Another 2D HF-He potential was reported by \citet{faj06} 
using the coupled-cluster method with single and double excitations with perturbative 
triple excitation (CCSD(T)). 
More recently, a three-dimensional (3D) PES was presented by \citet{sto03}.
This PES was computed using the Brueckner coupled-cluster method with
perturbative triples excitations (BCCD(T)) in the supermolecular approach, 
and  was fitted analytically using a kernel Hilbert space interpolation method.
This 3D potential was also used in close-coupling (CC) 
calculations of pure rotational excitation of HF in collisions with He by \citet{ree05}. 
 Cross sections for transitions for rotational levels up to $j=9$ 
 of HF were calculated for collision energies up to 2000 cm$^{-1}$.
Rate coefficients were obtained from 0.1 to 300 K. 
However, the PES of Stoecklin et al. predicts global and local minima with well
depths of of 43.70 and 25.88 cm$^{-1}$, respectively, compared to 39.20 and 35.12 cm$^{-1}$
for the experimentally derived PES of \citet{lov90}.

In this work, explicit quantum-mechanical close-coupling scattering calculations of
rotational quenching of HF in collisions with He to higher levels of rotational excitation were
carried out  using the
SAPT potential of \citet{mos94}.  The state-to-state rate coefficients are presented for a wide range of 
temperatures (0.1-3000 K), which will aid
in  modeling rotational spectra of HF in various astrophysical and atmospheric environments.
The computational method is discussed in Sect. 2, and 
we compare the PESs of Moszynski et al. (1994) and Stoecklin et al. (2003) in Sect. 3. The results are presented in Sect.
4, while astrophysical applications and scaling
approaches for estimating HF deexcitation by H$_2$ and H collisions are described in Sect. 5. 


\section{Quantum-scattering calculations}

We adopted the time-independent quantum-mechanical close-coupling (CC) theory developed  by \citet{art63} for 
the scattering of a linear rigid-rotor by an $S$-state atom. 
The state-to-state integral cross section for a transition from an initial rotational state  $j$ to
a final rotational state  $j'$ can be expressed as 
\begin{equation}
\sigma_{j\rightarrow j'}(E_{j})
=\frac{\pi}{(2j+1)k_{j}^2}\sum_{J=0}(2J+1)\sum_{l=|J-j|}^{J+j}
\sum_{l'=|J-j'|}^{J+j'}|\delta_{jj'}\delta_{ll'}
-S_{jj'll'}^J(E_j)|^2,
\label{eq_cross}
\end{equation}
where $\vec{j}$ and $\vec{l}$ are the rotational angular momentum of the HF molecule
and the orbital angular momentum of the collision complex, respectively. 
The total angular momentum $\vec{J}$ is given by $\vec{J}=\vec{l}+\vec{j}$. 
$S_{jj'll'}^J$ is an element of the scattering matrix, which is obtained by solving
coupled-channel equations and employing the usual  boundary conditions. 
$k_j=\sqrt{2\mu E_j}/\hbar$ denotes the wave vector    
for the initial channel,  $E_j$ is the kinetic energy for the 
initial channel,  and $\mu$ the reduced mass of the HF-He system.
The total quenching cross section from an initial state $j$ can be obtained 
by summing the state-to-state cross sections $\sigma_{j\rightarrow j'}(E_{j})$ over
all final $j'$ states,  where $j^{\prime} < j$.

The quantum-scattering code MOLSCAT \citep{molscat} was applied in 
the close-coupling calculations.
The propagation in $R$  was carried out to 50 \AA  \
with the coupled-channel equations solved using the modified
log-derivative Airy propagator \citep{ale87}.  
To ensure convergence in the cross-section calculations, 
at least five to ten closed channels in the basis and a sufficient
number of partial waves were included.   
 HF rotational energy levels are
given in Table 1, which were obtained using the rotational constant $B_e$=20.953 cm$^{-1}$
 \citep{iri07} and the centrifugal distortion constant $D$=0.0021199 cm$^{-1}$ \citep{cox90}.
The CC calculations were performed for collision energies ranging
from $10^{-5}$ to 15,000 cm$^{-1}$ with $\mu=3.3353$~u.

The rate coefficients for rotational transitions can be computed numerically
by thermally averaging the corresponding cross sections over a Maxwellian kinetic
energy distribution
\begin{equation}
k_{j\rightarrow j^{\prime}}(T) 
= \left (\frac{8}{\pi \mu \beta} \right )^{1/2}\beta^2\int^{\infty}_0 E_j 
\sigma_{j\rightarrow j^{\prime}}(E_j) \exp(-\beta E_j)dE_j,
\label{eq_rate}
\end{equation}
where $T$ is the temperature, $\beta=(k_BT)^{-1}$, and $k_B$ is Boltzmann's constant.

\section{Comparison of potential energy surfaces}

In the rigid-rotor scattering calculations,
the interaction potential of HF-He, $V(R,\theta)$, was expanded in terms of Legendre 
polynomials,
\begin{equation}
V(R,\theta)=\sum_{\lambda=0}^{\lambda_{\rm max}}v_{\lambda}(R)P_{\lambda}(\textrm{cos} \theta),
\end{equation}
\label{eq_vlm}
where $P_{\lambda}$ are Legendre polynomials of order $\lambda$, 
 $R$ is the distance between the HF center of mass and the He atom,
and  $\theta$  the angle between $\vec{R}$ and the HF molecular axis.
The angular dependence of the interaction potential was expanded to 
$\lambda_{\rm max}=20$.  

\begin{figure}
\advance\leftskip -0.0cm
\includegraphics[scale=0.6]{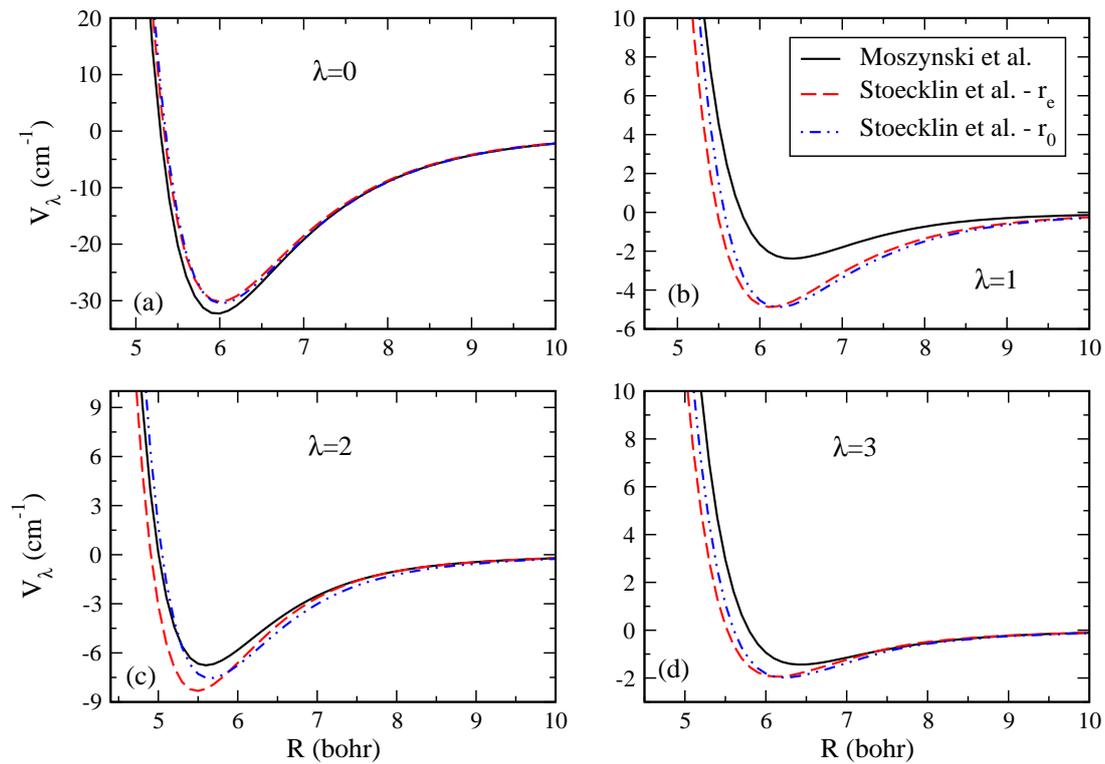}
\caption{
 Potential expansion terms $v_{\lambda}$, $\lambda$=0, 1, 2, and 3, for the HF-He PESs of Moszynski et al. (1994)
and Stoecklin et al. $r_e$ and $r_0$ are used to obtain 2D PESs for Stoecklin et al. (2003), while the   Moszynski et al.
PES was constructed for $r_e$.
}
\label{fig-lam}
\end{figure}

In Fig.~\ref{fig-lam}, the first four components of $v_{\lambda}(R)$, $\lambda$=0, 1, 2, and 3 
are plotted as a function of $R$ for the PESs of Moszynski et al. (1994) and Stoecklin et al. (2003).
To obtain a 2D rigid-rotor PES, the optimal approach is to average 
the 3D PES over the ground-state vibrational function of the 
diatomic molecule \citep{fau05,kal14}.
However, in this work, we mainly used the 2D PES of Moszynski et al. (1994),
which was obtained at $r_e$, to calculate the rotational
quenching cross sections and rate coefficients. For the 3D PES of Stoecklin et al., 
two values of the HF bond length, 
the equilibrium distance $r_e$=1.7328 a$_0$ and the vibrationally
averaged
bond distance for the ground vibrational state $r_0$=1.767 a$_0$ \citep{zhang},
were used to compute $v_{\lambda}(R)$. 
Comparing the plots, we can see that differences between the PESs of Moszynski et al.
and Stoecklin et al. are apparent, particularly for components $\lambda$=1 and 2. In the
case of the PES of Stoecklin et al., except for $\lambda=0$, some differences can be seen  
between the curves for $R$ less than 6 a$_0$ owing to the 
different values of the HF bond length. Therefore, one can expect discrepancies to arise in scattering calculations
performed on the two different PESs. We recall that the Moszynski et al. PES agrees very well with the experimentally deduced surface of \citet{lov90}.

\section{Results and discussion}

\subsection{State-to-state and total quenching cross sections}

\begin{figure}
\advance\leftskip -0.0cm
\includegraphics[scale=0.6]{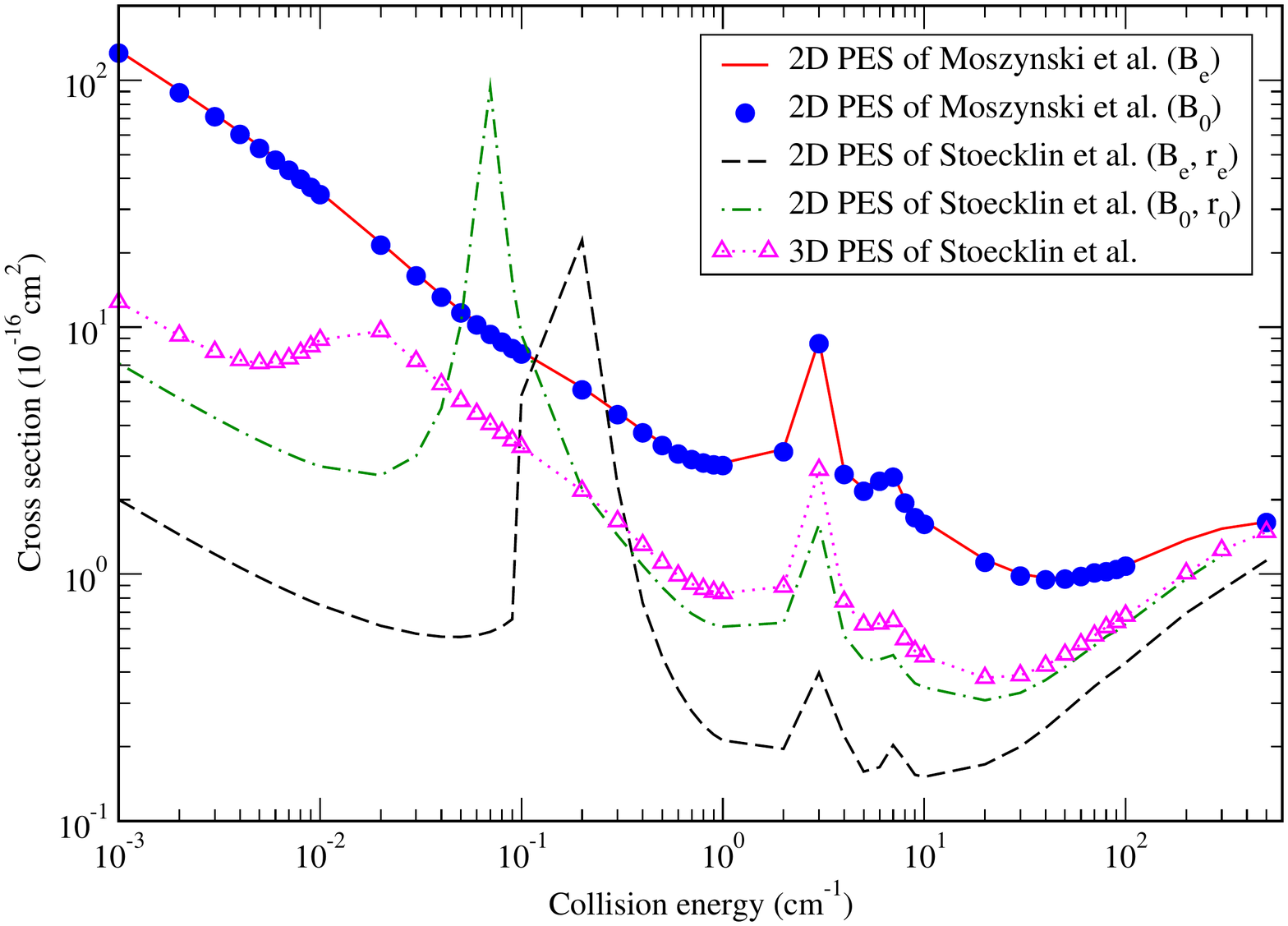}
\caption{
State-to-state rotational quenching cross sections from initial state $j=1$ of HF in 
 collisions with He.  
}
\label{fig-csj1}
\end{figure}

\begin{figure}
\advance\leftskip -1.0cm
\includegraphics[scale=0.8]{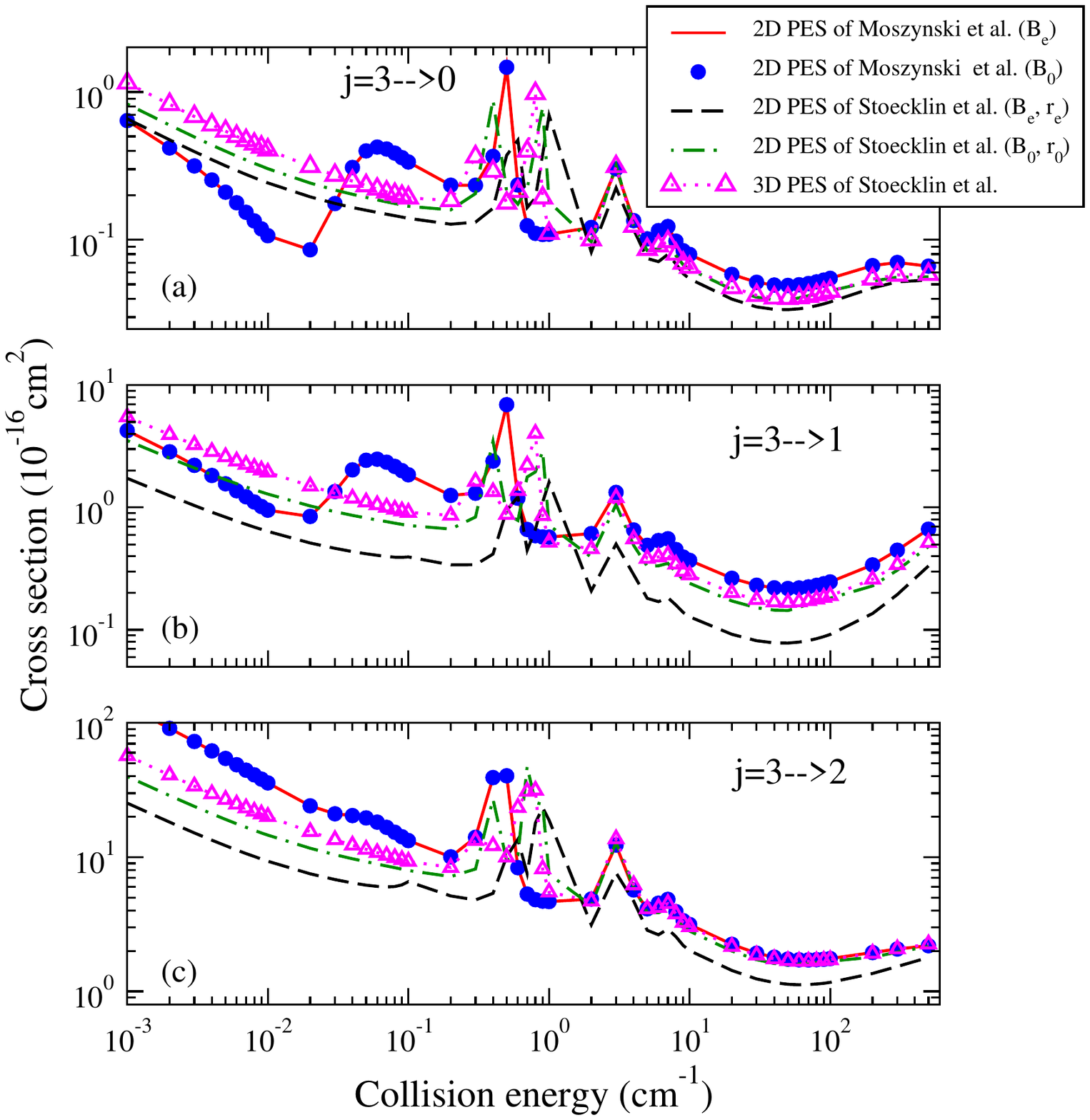}
\caption{
State-to-state rotational quenching cross sections from initial state $j=3$ of HF in 
 collisions with He.  
}
\label{fig-csj3}
\end{figure}

We performed calculations of the state-to-state quenching cross sections for initial HF
rotational states of  $j=1, \  2, \ \cdots, \ 20$ using the PES of Moszynski et al. 
(1994).\footnote{All state-to-state 
deexcitation cross sections and rate coefficients for HF-He 
are available on the UGA Excitation Database website
(www.physast.uga.edu/amdbs/excitation). The rate coefficients are also available in
the BASECOL \citep{dub13} 
and the Leiden Atomic and Molecular Database (LAMDA) \citep{sch05} formats.
In addition, estimates for HF-H$_2$ and HF-H rate coefficients obtained by the reduced-potential
scaling approach, described below, are included in the new LAMDA file along with the original
data from the LAMDA website.} \
To evaluate the accuracy of the current computed cross section and to compare with 
the results obtained using the PES of \citet{sto03},
calculations of state-to-state quenching cross sections from $j$=1 and 3 were also
performed using the PES of Stoecklin 
et al.\footnote{The rotational deexcitation cross sections displayed in Reese et al. (2005) were found to be discrepant
and not consistent with rate coefficients given in the same paper as confirmed by T. Stoecklin (priv. communication).}
In the rigid-rotor approximation calculations
carried out on the PES of Stoecklin et al.,
the 2D PESs are obtained by fixing the HF bond distance at $r_e$ and $r_0$.  
Correspondingly, rotational constants of HF $B_e$=20.9537 cm$^{-1}$ \citep{iri07} 
and $B_0=B_e-1/2\alpha_e$=20.5570 cm$^{-1}$ were used in the cross-section calculations, 
where the vibration-rotation interaction constant $\alpha_e=0.7934$ cm$^{-1}$ \citep{iri07}.
As examples, the state-to-state quenching cross sections from initial states  
$j$=1 and 3 are presented in Figs.~\ref{fig-csj1} and \ref{fig-csj3}, respectively.
In the case of quenching $j=1\rightarrow 0$, Fig.~\ref{fig-csj1} illustrates that 
there are significant differences between the cross sections obtained using 
the PESs of Moszynski et al. and Stoecklin et al. 
that are due to the different structures of the two PESs, as shown in Fig.1. 
For the PES of Stoecklin et al.,
the cross sections calculated using ($B_0$, $r_0$) agree better with the results
obtained using their 3D potential. 
A number of resonances in the cross section, which occur for low collision energies associated with
the van der Waals wells, demonstrate their sensitivity to the adopted PES.

The PES of Moszynski et al., used in our calculations, was constructed by 
fixing the HF bond distance at its equilibrium value $r_e$. We used the rotational
constant $B_e$ to evaluate the HF rotational energy levels (see Table~\ref{table1}). 
However, to study the effect of the rotational constant on the cross sections, we carried out 
cross-section calculations
from initial states $j=1$, 3, and 10 using the rotational constant $B_0$. 
Figures~\ref{fig-csj1} and \ref{fig-csj3} show that for the PES of Moszynski et al.,  
the state-to-state quenching cross sections from initial $j=1$ and 3 using $B_e$ and $B_0$
are nearly identical. For state-to-state quenching cross sections from initial $j=10$ (not shown), 
the differences between the results obtained using $B_e$ and $B_0$ are generally lower than 10\%. 

In Fig.~\ref{fig-csj3} we compare the state-to-state quenching cross section from initial state
$j=3$. As observed in the case of $j=1$, the cross sections display resonances 
in the intermediate energy region from $\sim$0.01 cm$^{-1}$ to $\sim$10 cm$^{-1}$ due to 
quasibound levels supported by the attractive part of the interaction potential. 
 Importantly, for astrophysical applications, 
the properties of the resonances influence the quenching rate coefficients
at low temperatures.  
In contrast to initial $j=1$, the difference between the cross sections obtained on the
two PESs is smaller. In particular, for the PES of \citet{sto03} the results using 
($B_0$, $r_0$) are similar to the results using the 3D potential for collision energies higher
than 1.0 cm$^{-1}$.
As can be seen, the deexcitation process from $j=3$ is dominated by the $\Delta j=-1$ transition.
Furthermore, the computed cross sections show that 
the  $\Delta j=j'-j = -1$  transition dominates the 
deexcitation for all $j$, and the cross sections generally increase
with increasing $j'$ , with that for $j'=0$ being the smallest.

\begin{figure}
\advance\leftskip -0.0cm
\includegraphics[scale=0.6]{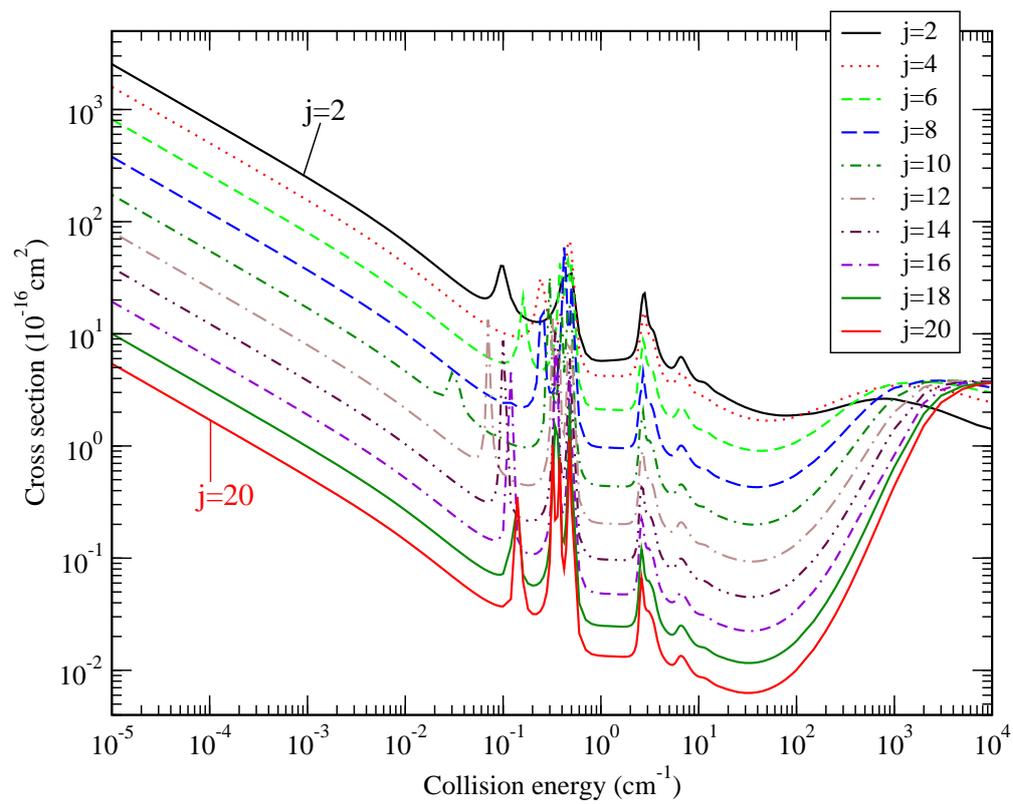}
\caption{
Total deexcitation cross sections from initial states $j$=2, 4, 6, 8, 10, 12, 14, 16, 18, 
and 20 of HF in collisions with He with the PES of Moszynski et al. (1994). 
}
\label{fig-totcs}
\end{figure}

The total quenching cross section from a given initial level $j$ can be computed 
by summing over all final states $j'$.
In Fig.~\ref{fig-totcs} the total quenching cross sections from
selected initial levels $j$=2, 4, 6, $\cdots$, 18, and 20 are displayed.  
Generally, the total quenching cross sections 
have similar behavior, decreasing with $j$ for $E_j\leq$ 50 cm$^{-1}$,
 but differences result for small $j$ at high energy that are
due to a
limited number of final exit channels. 
Each of the cross sections 
exhibit the behavior predicted by the Wigner (1948) threshold law
 at ultra-low collision energies below $\sim$10$^{-3}$ cm$^{-1}$, 
where only $s$-wave scattering
contributes and the cross sections vary inversely with the relative velocity. 
In the intermediate energy region, between 0.1 and 
10 cm$^{-1}$, the cross sections display scattering resonances,
but they reveal somewhat different structures depending on
the initial rotational state $j$ between 0.05 and 1 cm$^{-1}$.
Except for $j$=2, the total deexcitation cross sections decrease 
to a global minimum near 50 cm$^{-1}$.

\subsection{State-to-state quenching rate coefficients}

\begin{figure}
\advance\leftskip -0.0cm
\includegraphics[scale=0.6]{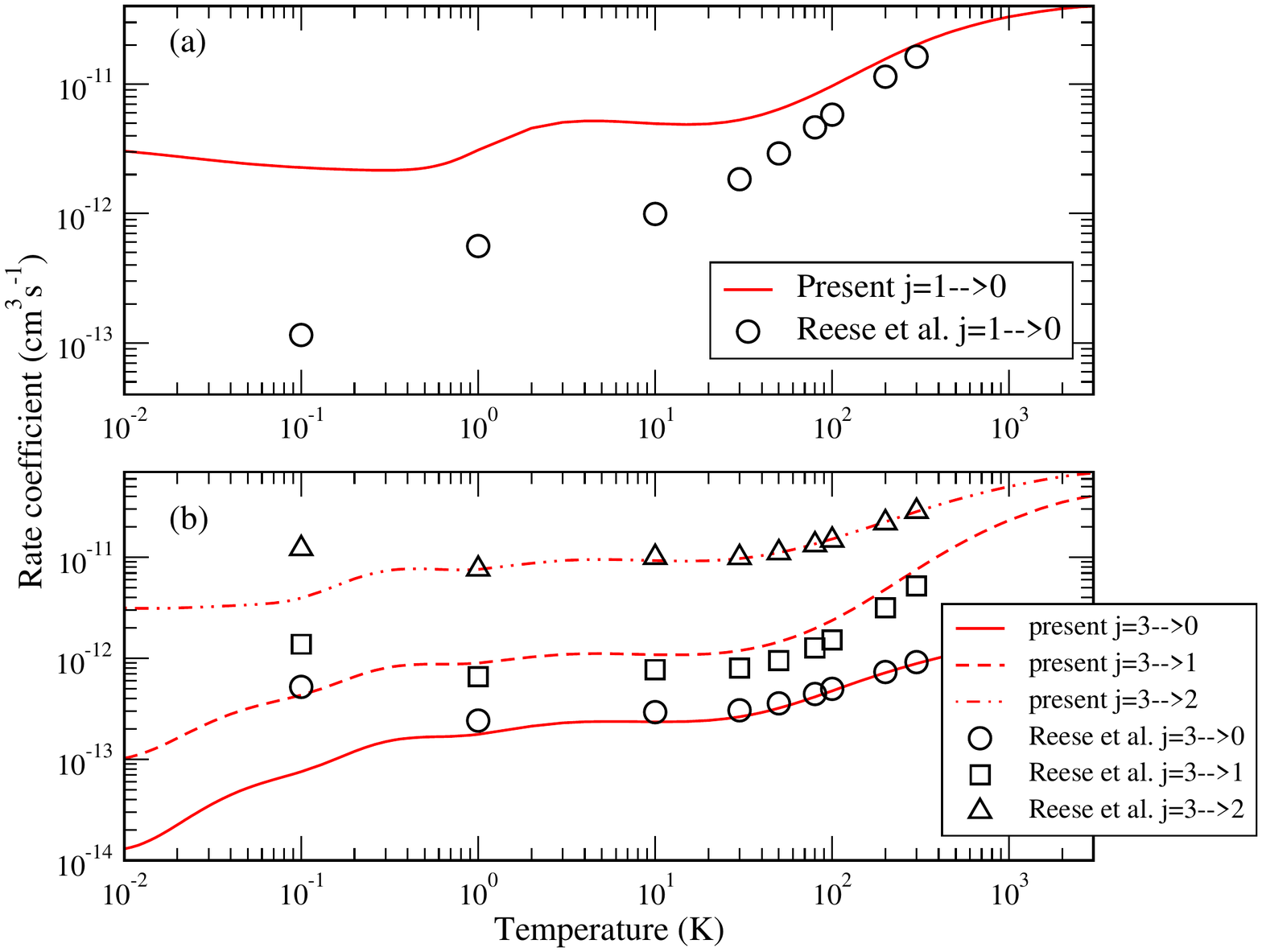}
\caption{ 
State-to-state rotational quenching rate coefficients from initial states 
$j$=1 and 3 of HF that are due to collisions with He. (a) $j=1$, (b) $j=3$.
Lines:  present results with the PES of Moszynski et al. (1994); 
symbols: \cite{ree05} with the PES of Stoecklin et al. (2003). 
}
\label{fig-ratej13}
\end{figure}

\begin{figure}
\advance\leftskip 0.5cm
\includegraphics[scale=0.6]{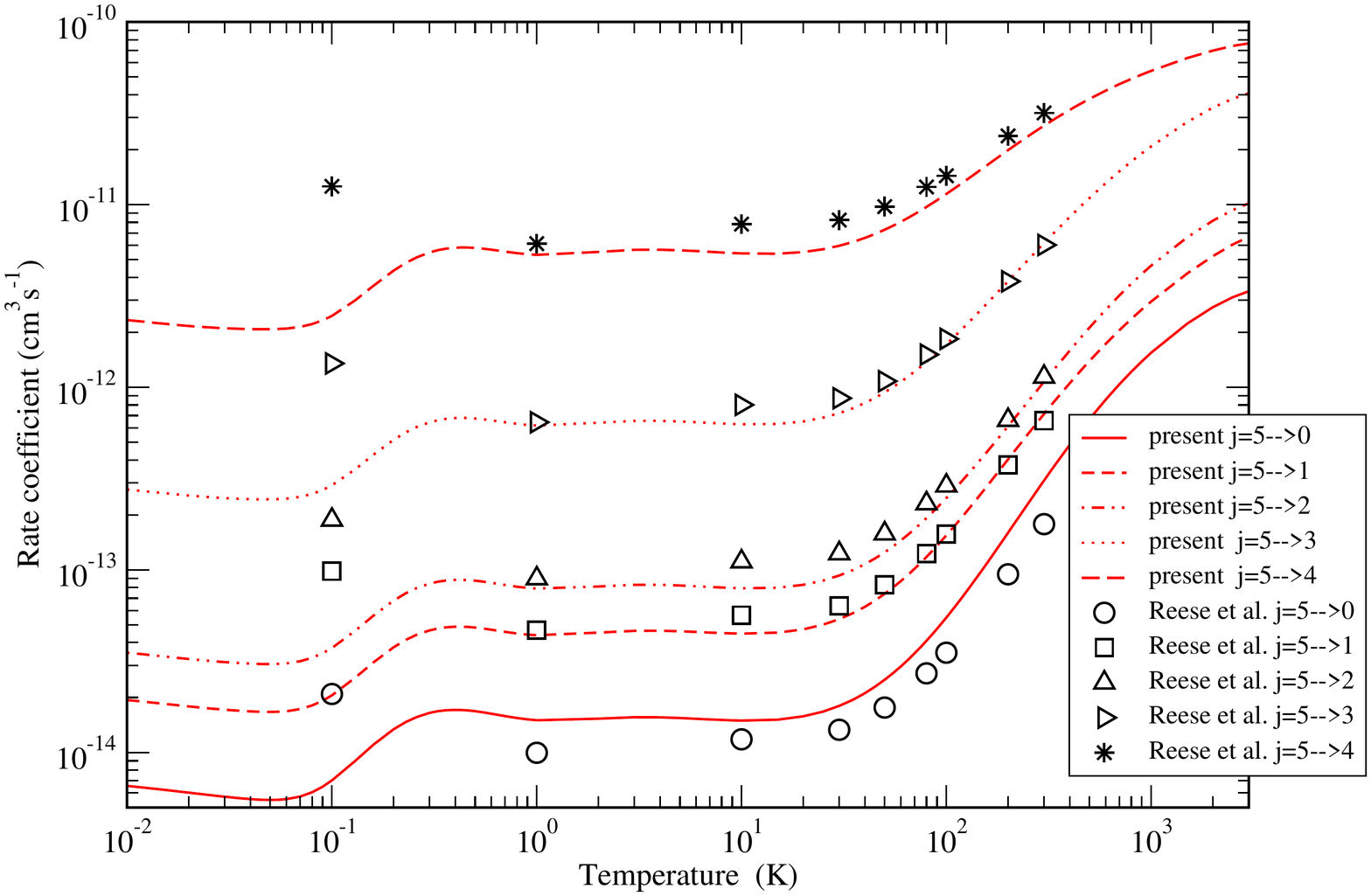}
\caption{ 
State-to-state rotational quenching rate coefficients from initial states 
$j$=5 of HF that are due to collisions with He. 
Lines:  present results with the PES of Moszynski et al. (1994); 
symbols: \cite{ree05} with the PES of Stoecklin et al. (2003). 
}
\label{fig-ratej5}
\end{figure}

\begin{figure}
\advance\leftskip -0.0cm
\includegraphics[scale=0.6]{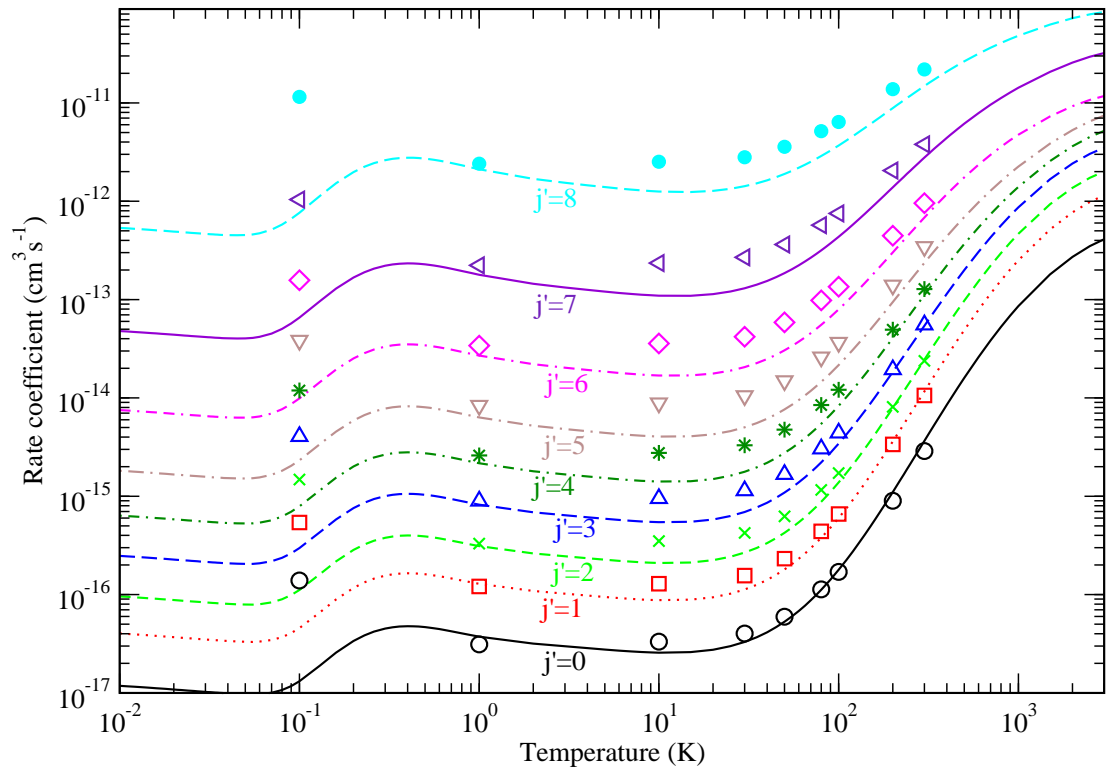}
\caption{ Same as Fig.~\ref{fig-ratej5}, but for initial rotational state $j=9$.}
\label{fig-ratej9}
\end{figure}

The quenching rate coefficients can be computed by averaging the appropriate cross 
sections over a Maxwell-Boltzmann distribution of collision energy $E_j$, as given by
Eq.~\ref{eq_rate}. 
The state-to-state quenching rate coefficients for initial HF
rotational states of  $j=1, \  2, \ \cdots, \ 20$ were
calculated using $B_e$ and the PES of Moszynski et al. (1994).
However, to our knowledge, there are no published experimental rate coefficients available. 
Our rate coefficients, computed using the PES of Moszynski et al., 
are only compared with the theoretical results of \citet{ree05} ,
which were obtained over the limited temperature range of 0.1 to 300~K.
As examples, Figs.~\ref{fig-ratej13}, \ref{fig-ratej5}, and \ref{fig-ratej9} 
present selected  state-to-state quenching rate coefficients from 
initial rotational levels $j$=1, 3, 5, and 9.  Except for deexcitation
from $j=1$, the current results generally 
follow similar trends with that of Reese et al., which were computed 
using the 3D potential of \citet{sto03}. However, at 0.1 K, the rate coefficients of
Reese et al. are always significantly larger than the current results for all 
transitions, except for $j=1$. 

\begin{figure}
\advance\leftskip -0.0cm
\includegraphics[scale=0.6]{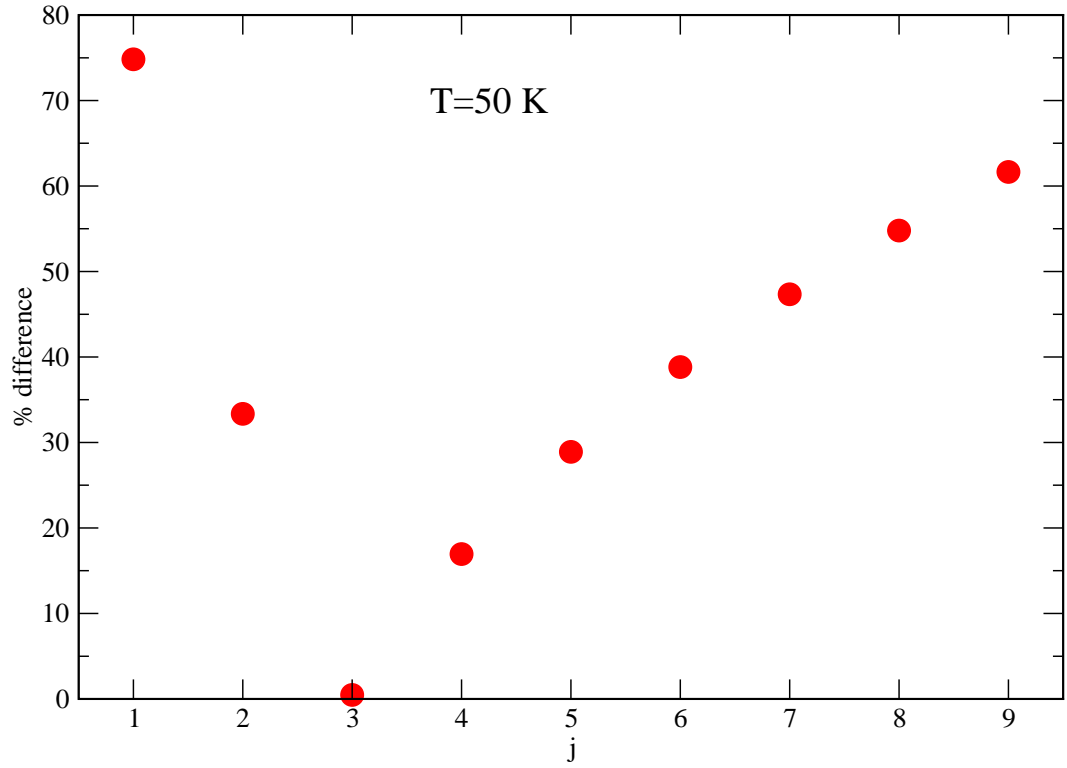}
\caption{ 
Percent difference of HF-He rate coefficients for the deexcitation of level $j$ for the dominant transition $\Delta j=-1$ at 50 K 
between current results using the PES of Moszynski et al. (1994) and
the results of Reese et al. (2005) with the PES of Stoecklin et al. (2003). 
}
\label{fig-diff}
\end{figure}

For initial state $j=1$, Fig.~\ref{fig-ratej13}(a) shows that the 
current rate coefficients are larger than the results of Reese et al.
As for initial state $j=3$ shown in Fig.~\ref{fig-ratej13}(b), for temperatures above 1 K,
the current state-to-state rate coefficients agree reasonably
well with
that of Reese et al., although our results are somewhat larger
for the $j=3\rightarrow 1$ transition.
State-to-state quenching rate coefficients from  initial states $j=5$ and 9 
are compared in Figs.~\ref{fig-ratej5} and \ref{fig-ratej9}, respectively, with the results 
of Reese et al. Except for the deexcitation to the final state $j'=0$ and for 0.1 K, 
the current rate coefficients are smaller than those of Reese et al., 
similar to what is found for $j=3$. 
For each dominant deexcitation transition, $\Delta j=j'-j= -1$,
$j$=1, 2, $\cdots$, 9, at a temperature of 50 K, we compare the percent
differences between our 
rate coefficients and the results of Reese et al. (2005). 
In Fig.~\ref{fig-diff}, the percent differences are displayed as a function of initial rotational
state $j$. The percent difference is near zero for $j$=3, but$\text{}$  
the differences vary from 20\% to 75\% for all other \textit{j
\textit{ \textup{values}}}. 

\begin{figure}
\advance\leftskip -0.0cm
\includegraphics[scale=0.6]{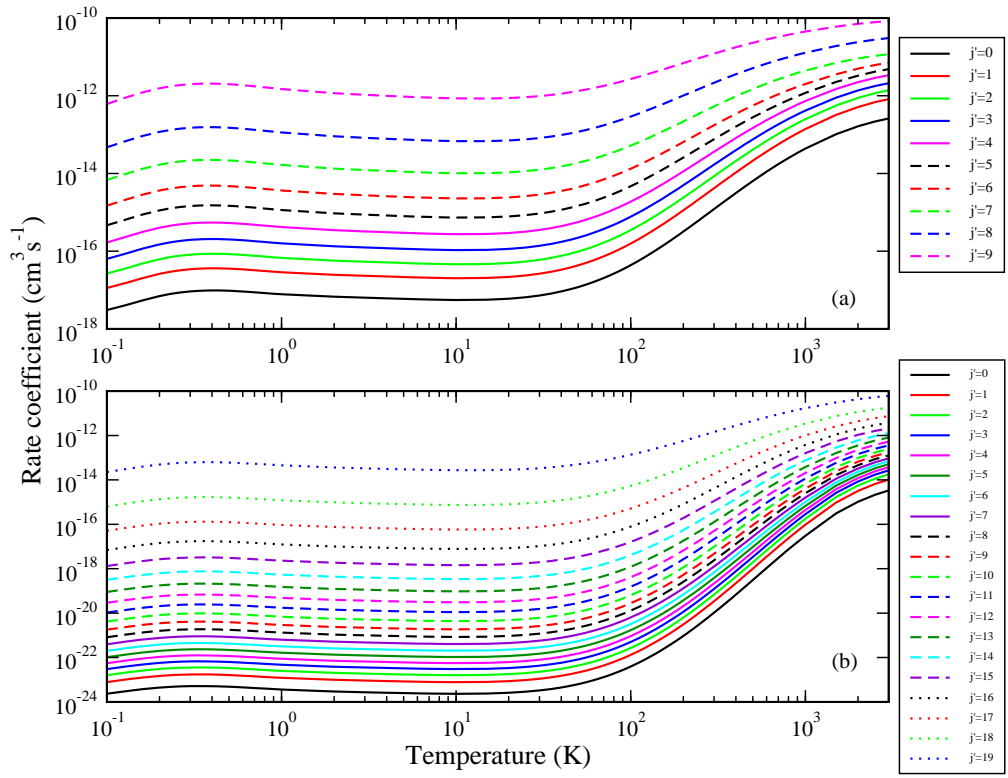}
\caption{ 
State-to-state rotational deexcitation rate coefficients from initial state $j=10$ 
and $j=20$ of HF in collisions with He using the PES of Moszynski et al. (1994). 
(a) $j=10$, (b) $j=20$.
}
\label{fig-ratej10-20}
\end{figure}

For illustration, in Fig.~\ref{fig-ratej10-20} we present the 
total deexcitation rate coefficients
for temperatures ranging from 0.1 K to 3000 K for initial states $j$=10 and 20. 
Over the whole temperature range considered, the rate coefficients generally
increase with increasing temperature for all transitions.
Furthermore, the rate coefficients clearly decrease with increasing 
$|\Delta j|=|j'-j|$ with the $\Delta j=-1$ transitions dominant.

\section{Applications}

As highlighted in the Introduction, HF has been observed in both emission and absorption in a variety
of astronomical environments that may be characterized by diverse physical conditions. As a consequence,
the rotational levels of HF may be populated by different, or multiple, mechanisms leading to spectra differing from LTE.
For the UV-irradiated environment of the Orion Bar,  \citet{van12b} considered electrons and H$_2$ as possible
impactors for inelastic collisional excitation. However, given that the Orion Bar is a prototypical photodissociation
region (PDR), the abundances of both H and He are most likely higher than or comparable to that of H$_2$. 
In many PDR enviroments,  collisions due to all four colliders may need to be considered.

For a variety of reasons, it is often the case that excitation rate coefficients for a molecular species may only
be available for He collisions, as performed in the current work. In such instances, a common practice is to estimate H$_2$, and occasionally H, collisional
excitation rate coefficients using He data by scaling by the square root of the ratio of the collision systems' reduced masses, here a factor
of 1.4 for H$_2$. \citet{wal14} demonstrated on both theoretical and numerical grounds that this {\it \textup{standard}} reduced-mass scaling
approach is typically invalid. We therfore do not recommend here that such an approach be adopted with the present HF-He rate
coefficients. Fortunately, explicit HF-H$_2$ rate coefficients for some rotational quenching transitions of the HF ground vibrational state have
been computed by \citet{gui12}.

\subsection{Prediction of HF rate coefficients by scaling}

\begin{figure}
\advance\leftskip 1.0cm
\includegraphics[scale=1.0]{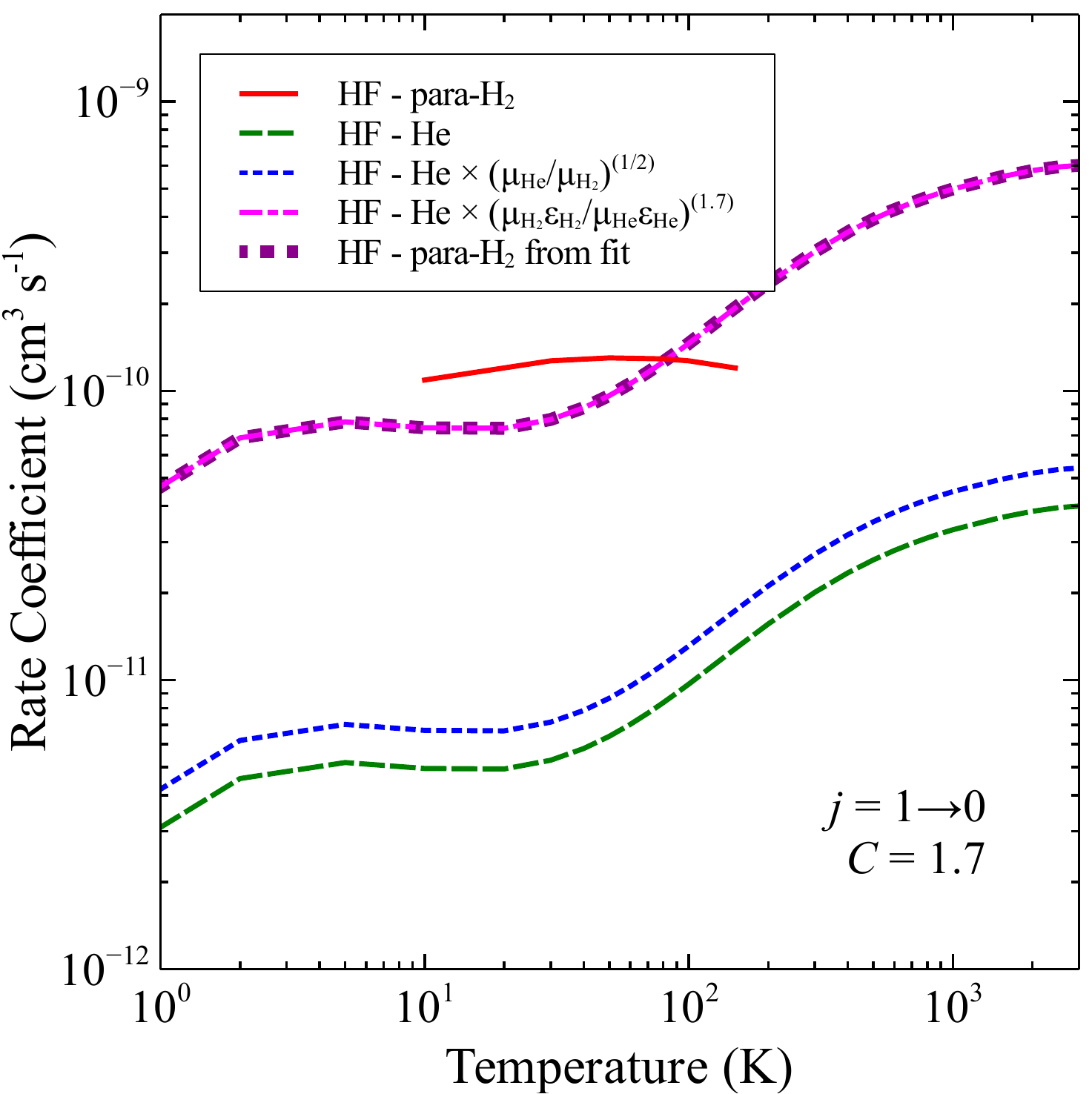}
\caption{ 
HF--para-H$_2$ \citep{gui12} and HF-He rate coefficients (current results) compared
to  HF--para-H$_2$
rate coefficients deduced by  various scaling approaches for the $j=1\rightarrow 0$
transition. See text for discussion.
}
\label{scale10}
\end{figure}

Given the availability of HF-H$_2$ and HF-He collisional data, we can test various scaling
methods, including the reduced-potential approach introduced in \citet{wal14}.
In the reduced-potential method, the collisional
data are scaled by the reduced potentials $\mu_{X}\varepsilon_{X}$
 according to
\begin{equation} \label{scale}
k^{\rm Z}_{j\rightarrow j'}(T) =
\left (\frac{\mu_{\rm Z}\varepsilon_{\rm Z}}{\mu_{\rm Y}\varepsilon_{\rm Y}}\right)^{C}
k^{\rm Y}_{j\rightarrow j'}(T),
\end{equation}
 where $\mu_{}$ is the
 reduced mass of the HF-$X$ system, $\varepsilon_{X}$ is the van der Waals
 minimum of the HF-$X$ system, and $C$ is a phenomenological exponent.
 $X=Y$ is usually He, with $X=Z$ typically para-H$_2$.
The reduced-potential and standard reduced-mass approaches are compared in Fig.~\ref{scale10} for the HF rotational deexcitation transition $j=1\rightarrow 0$. Standard reduced-mass
scaling results in an estimate for para-H$_2$ collisions that is a factor of 10 too small, while the reduced-potential method with an exponent of $C=1.7$ agrees well. 

Given its better performance, the reduced-potential scaling approach of \citet{wal14}
 was therefore used to predict unknown rate coefficients for para-H$_2$ and H colliders with HF. We adopted 
 the He collision data computed here and the smaller set of
 para-H$_2$ rate coefficients for $j<5$ calculated by
 \citet{gui12} for $T=10-150$~K.  The
 rate coefficients are very sensitive to the presence of quasibound resonances over
 this temperature range, which may partially be accounted for with the reduced-potential
 approach since it takes into the account the different interaction well-depths.
Compared to the reduced-mass scaling technique,
 where the rate coefficients simply scale as the square root
 of the ratio of the reduced masses, the reduced-potential scaling approach
 offers an improvement for predicted HF rate coefficients, much like the improved CO and
 H$_2$O rate coefficient predictions of \citet{wal14}.
 While the value of $C$ for CO rate coefficients ranged from $-0.2$  to 1.3 and
 from 0.5 to 1.2 for H$_2$O, it was noted that symmetries were involved in
 obtaining the best value of $C$.

\begin{figure}
\advance\leftskip 1.0cm
\includegraphics[scale=0.8]{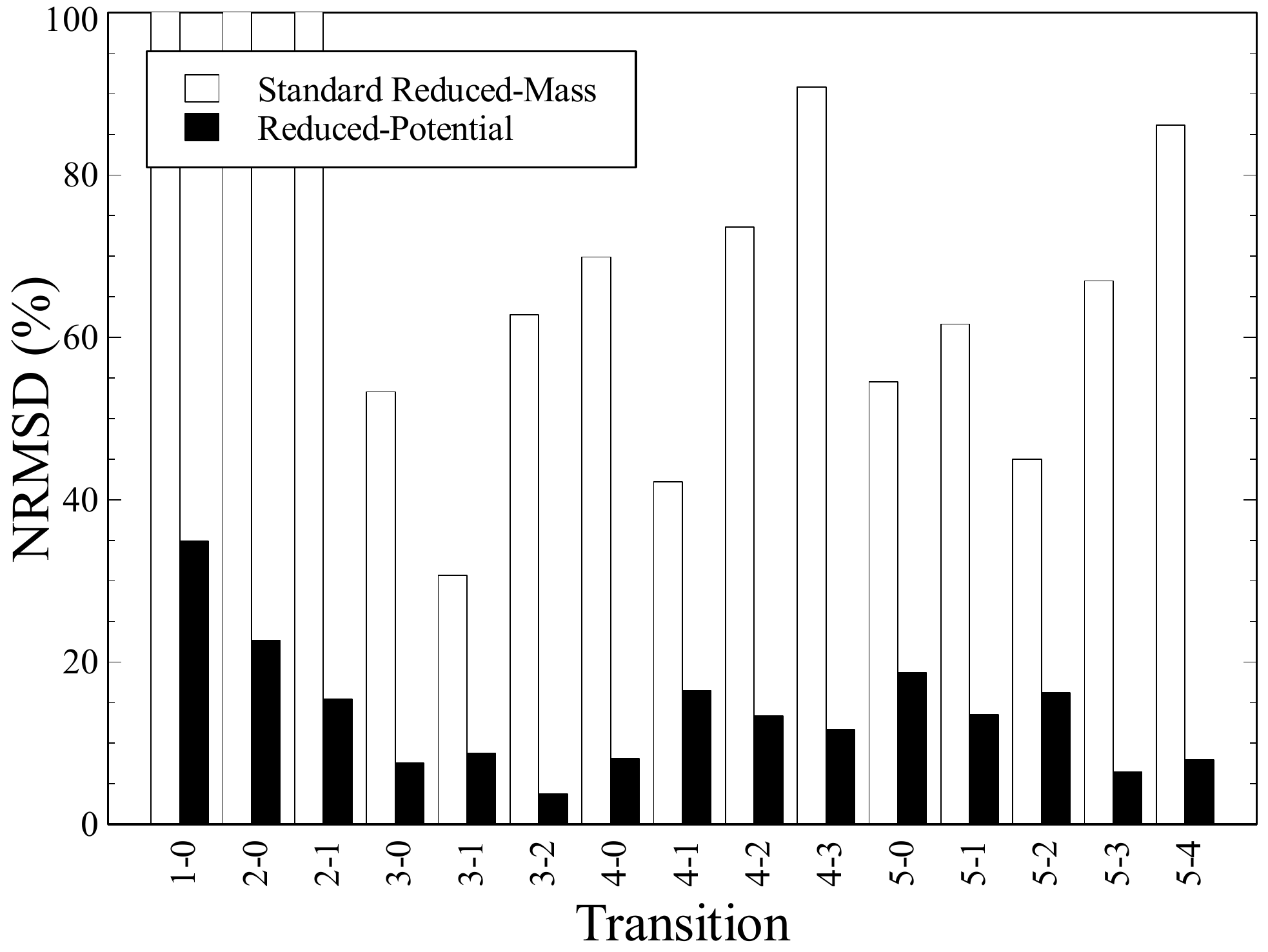}
\caption{
Normalized root-mean-square deviation (NRMSD) in standard reduced-mass
scaling and reduced-potential scaling for 15 transitions of HF-para-H$_2$, truncated
at 100$\%$.
\label{fighist}}
\end{figure}

\begin{figure}
\advance\leftskip 1.0cm
\includegraphics[scale=1.0]{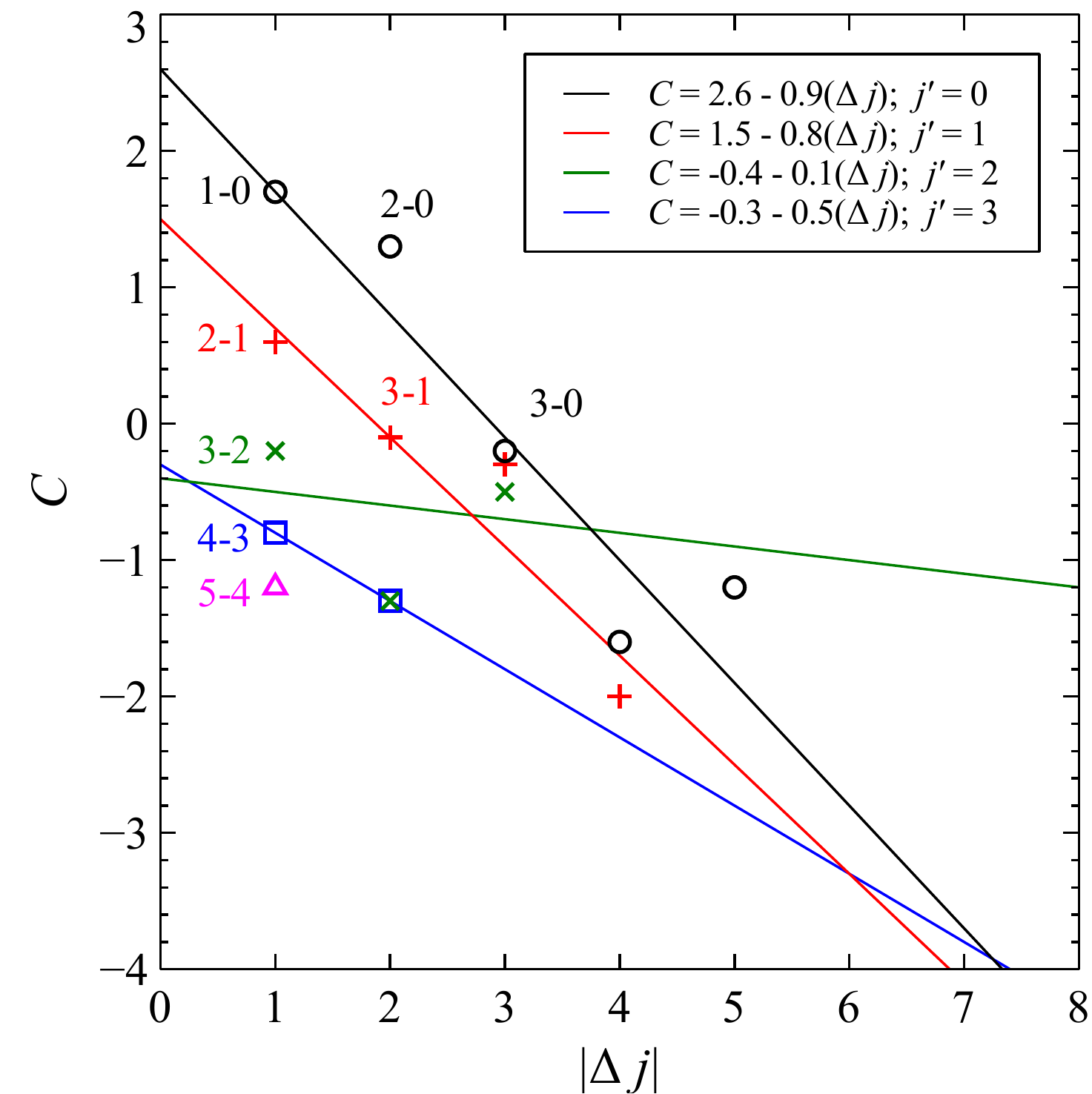}
\caption{
Phenomological constant $C$ as a function of $|\Delta j|$. A linear
least-squares analysis was performed for each $j'$ and the resulting
linear functions are plotted. Note the convergence of these functions
(except for $j'=2$) around $|\Delta j|=6$ and $C=-3$. Symbols correspond
to $C$ determined for actual HF-para-H$_2$ and HF-He data.
\label{fig-C}}
\end{figure}

 The accuracy of the reduced-potential scaling approach can be further tested here with
 the heteronuclear molecule HF.
 The normalized root-mean-square deviation (NRMSD), $\sigma_{norm}$,
 quantifies the residual variance between the calculated
 H$_2$ rate coefficients, $k_{calc}$, and those scaled from He, $k_{scale}$,
 and is given by
\begin{equation} \label{nrmsd}
\sigma_{norm} =
\frac{\sqrt{\frac{\sum\limits_{T=i}^{N} (k_{scale}(T)-k_{calc}(T))^2}{N}}}
{k_{max} - k_{min}},
\end{equation}
 where $N$ is the number of temperature data points and
 $k_{max}$ and $k_{min}$ are the values of the maximum and
 minimum rate coefficients.
 The resulting NRMSD percentages for the H$_2$ rate coefficient predictions
 for both standard reduced-mass scaling and reduced-potential scaling is
 given in Table~\ref{table2} and Fig.~\ref{fighist} for the first fifteen transitions of HF
 at 50~K.
 When comparing the NRMSD for both methods for each transition of HF, the
 reduced-potential scaling predictions exhibit less residual variance
 in all fifteen transitions with a mean of 14$\%$ in its 
ability to reproduce the explicit HF-H$_2$ calculations as computed by \citet{gui12}. In many cases, the NRMSD for standard
reduced-mass scaling exceeds 100\%, while the reduced-potential approach gives NRMSD $<35$\%.
A linear least-squares analysis was then performed, using the first fifteen
transitions of HF, for each $j'$ and the resulting
 linear functions are plotted in Fig.~\ref{fig-C}.
 Except for $j'=2$, the lines converge around $|\Delta j|=6$ and $C=-3$.
 For $|\Delta j|=1$, the values of $C$ decrease linearly with $j'$ and
 are listed in Table~\ref{table3} for $j^{\prime}=0-4$.
The optimal values for $C$, valid for all temperatures, were forced to exactly reproduce the reduced-potential scaling result for
 the dominant $\Delta j = -1$ transitions.
 
\begin{figure}
\advance\leftskip 1.0cm
\includegraphics[scale=1.0]{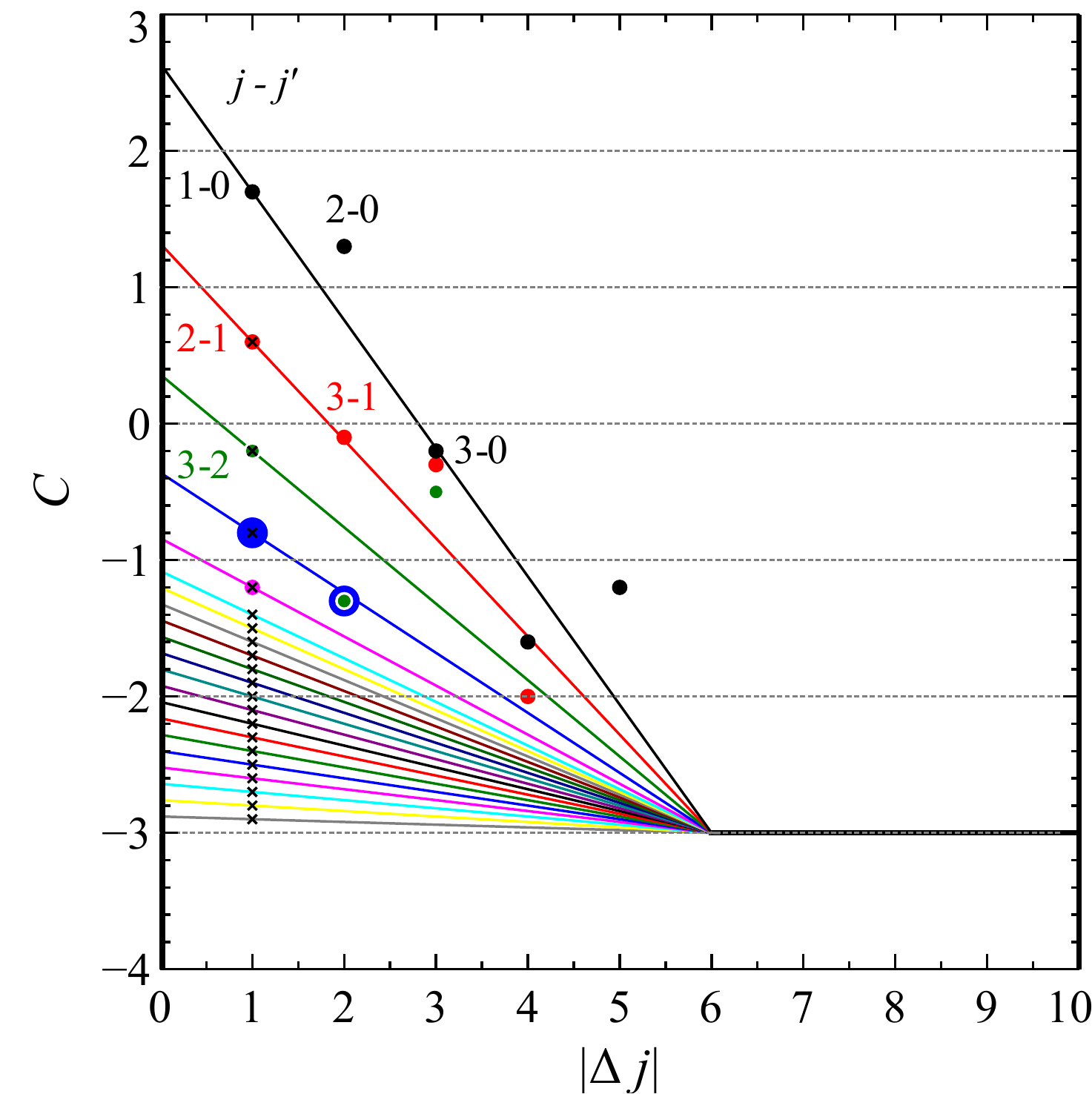}
\caption{
Phenomological constant $C$ as a function of $|\Delta j|$. As
a result of the linear decrease of $C$ with $j'$ and the convergence around
$|\Delta j|=6$ and $C=-3$, the slope and $y$-intercept can be obtained for
each transition and the value of $C$ can be predicted.
Symbols and fit lines are for $j^{\prime}=0-4$, same as in Fig.~12.
\label{figC-predict}}
\end{figure}

\begin{figure}
\advance\leftskip 1.0cm
\includegraphics[scale=1.0]{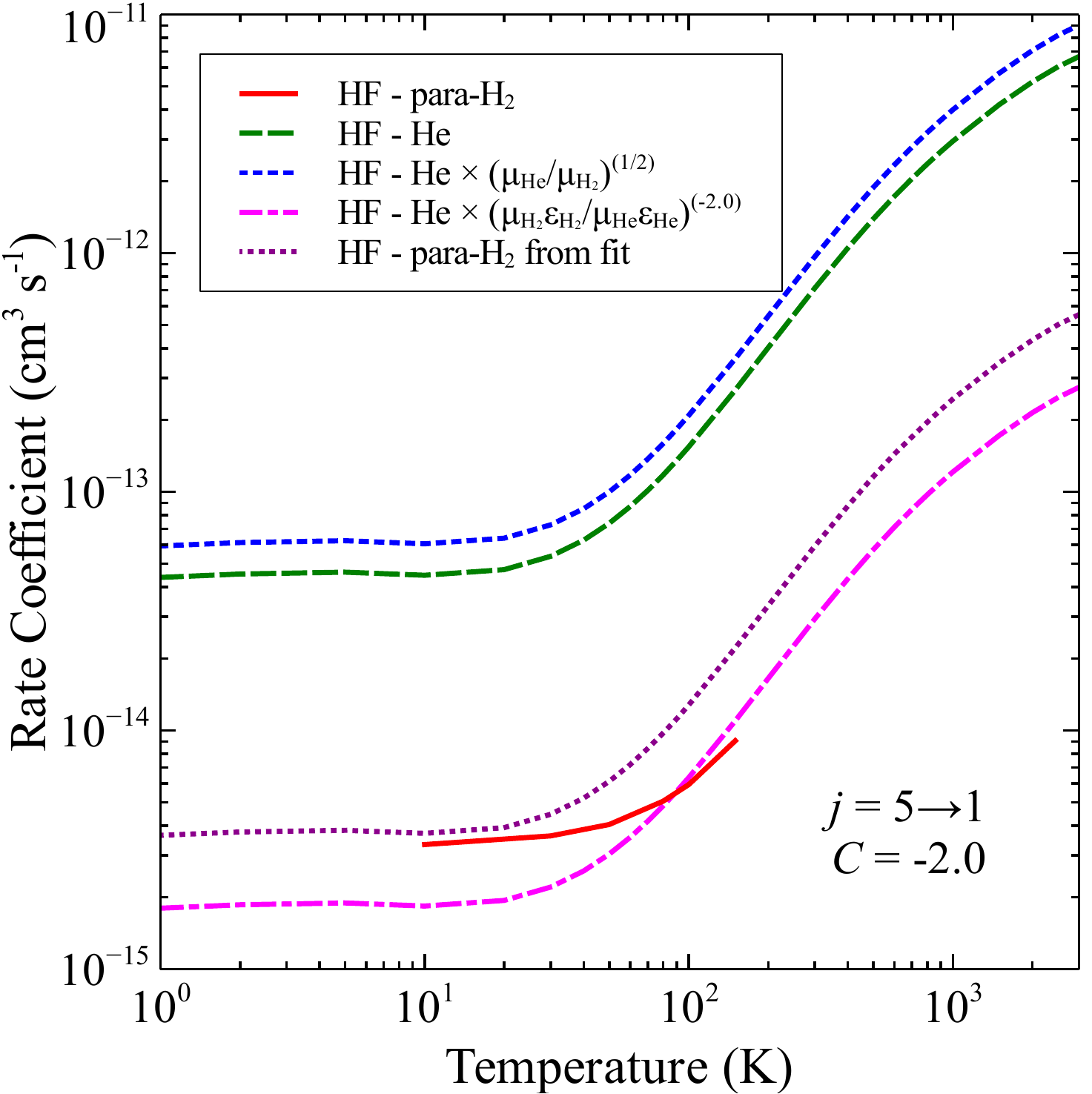}
\caption{ 
Same as Fig. 10, but for the $j=5\rightarrow 1$ transition.
}
\label{scale51}
\end{figure}

To use the reduced-potential approach in astrophysical applications, however, one needs estimates of $C$
when data for the impactor of interest $Z$, e.g. para-H$_2$, are unknown.
 Using the information from Fig.~\ref{fig-C}, the slope and $y$-intercept can be obtained for
each transition and the value of $C$ predicted as given in Fig.~\ref{figC-predict} and Table 3
for $j^{\prime} = 5-20$. The best prediction for transitions with $|\Delta j| \geq 6$
 is obtained with $C=-3$.
 Rate coefficients for the transitions of H$_2$ with $j<5$ are
  reproduced reasonably well by the reduced-potential approach with
 $C$ decreasing with increasing $|\Delta j|$ and increasing $j'$.
 Figures~\ref{scale10} and \ref{scale51} compare rate coefficient estimates using the fit prediction of $C$ to direct
 reduced-potential and reduced-mass values for select transitions.
 Weak transitions are scaled with $C=-3$, and since these rate coefficients
 are several orders of magnitude smaller than the dominant transitions,
 larger error in the predicted values is acceptable.

Given that the available HF-H$_2$ rate coefficients are limited to $j\leq 5$ \citep{gui12}, we used the reduced-potential scaling
method predictions for $C$ and the current HF-He data to estimate HF rotational deexcitation for $j=6-20$ for para-H$_2$ collisions. 
Furthermore, as there is a complete lack of HF rotational excitation
data for H impact, we extended the reduced-potential scaling approach to estimate HF-H deexcitation rate coefficients from
the current HF-He rate coefficients with the trend in $C$ taken from Fig.~\ref{figC-predict}. In using the reduced-potential approach, the following
parameters were adopted for HF collisions with He, H$_2$, and H: $\mu_{\rm He}=3.3353$ u, $\mu_{\rm H_2}=1.818$ u, $\mu_{\rm H}=0.9596$ u, 
$\epsilon_{\rm He}=39.68$ cm$^{-1}$ \citep{mos94}, $\epsilon_{\rm H_2}=359.0$ cm$^{-1}$ \citep{gui08}, and $\epsilon_{\rm H}=100.0$ cm$^{-1}$ \citep{sta96}. 
All rate coefficient data are available in the LAMDA format, as mentioned in footnote 1.
The availability of a complete set of HF rotational quenching rate coefficients due to collisions with H$_2$, He, H, and e$^-$ will allow for
detailed modeling of HF rotational emission lines in a variety of environments with a varying molecular fraction. However, we caution that the data obtained via reduced-potential 
scaling are approximate, but are reasonable estimates until explicit calculations become
available.

\section{Conclusion}

Rate coefficients for the deexcitation of rotational excited HF due to He collisions were computed using the close-coupling method and
an accurate potential energy surface. The adopted ab initio PES agrees well with an experimentally deduced empirical PES.
Differences of 75\% and larger were found with previous HF-He scattering calculations that used a less reliable PES. New rate coefficients
were obtained for HF rotational levels $j=1-20$ for 0.1-3000 K due to He. A recently introduced scaling approach was used to estimate rate
coefficients for HF-H  and missing HF-H$_2$  collisional data.

\vskip 1.0 cm

\begin{acknowledgements}
Work at UGA was supported by NASA grant NNX12AF42G, at Penn State by NSF Grant No. PHY-1203228, 
and at UNLV by NSF Grant No. PHY-1205838.  We thank Ad van der Avoird and T. Stoecklin for sending
their potential subroutines and for helpful correspondence.
\end{acknowledgements}

\newpage

\begin{table}
\begin{center}
\caption{Computed rotational excitation energies (cm$^{-1}$) of HF.
\label{table1}}
\vskip 0.5cm
\begin{tabular}{c c @{\hspace{1.0cm}} c c }
\hline \hline
  {$j$} &       {$E_j$}     &  {$j$} &       {$E_j$}     \\ [0.5ex]
  \hline
    0   &         0.00000   &   16   &      5540.34896   \\   [1ex]
    1   &        41.89886   &   17   &      6210.52398   \\   [1ex]
    2   &       125.64498   &   18   &      6914.70306   \\   [1ex]
    3   &       251.13516   &   19   &      7651.95740   \\   [1ex]
    4   &       418.21460   &   20   &      8421.30660   \\   [1ex]
    5   &       626.67690   &   21   &      9221.71866   \\   [1ex]
    6   &       876.26406   &   22   &     10052.10998   \\   [1ex]
    7   &      1166.66648   &   23   &     10911.34536   \\   [1ex]
    8   &      1497.52296   &   24   &     11798.23800   \\   [1ex]
    9   &      1868.42070   &   25   &     12711.54950   \\   [1ex]
   10   &      2278.89530   &   26   &     13649.98986   \\   [1ex]
   11   &      2728.43076   &   27   &     14612.21748   \\   [1ex]
   12   &      3216.45948   &   28   &     15596.83916   \\   [1ex]
   13   &      3742.36226   &   29   &     16602.41010   \\   [1ex]
   14   &      4305.46830   &   30   &     17627.43390   \\   [1ex]
   15   &      4905.05520   &        &                   \\   [1ex]
\hline\hline
\end{tabular}
 \tablefoot{$E_j = B_ej(j+1)-Dj^2(j+1)^2$, where $B_e$ and $D$ are the rotational constant  
  and the centrifugal distortion constant of HF, respectively.}
\end{center}
\label{tab1}
\end{table}

\begin{table}
\begin{center}
\caption{Optimized values of $C$ and their respective normalized
root-mean-square deviations (NRMSD) for collisional deexcitation transitions
of HF with H$_{2}$ scaled via the standard reduced-mass (rm) and
reduced-potential (rp) methods from HF-He collisional rate coefficients.
\label{table2}}
\vskip 0.5cm
\begin{tabular}{c c c c }
\hline \hline
$j\rightarrow j'$   & C & NRMSD$_{\rm rm}$   & NRMSD$_{\rm rp}$ \\
\hline
1-0     &       1.7     &       1058.85 &       34.90   \\
2-0     &       1.3     &       299.75  &       22.66   \\
2-1     &       0.6     &       161.28  &       15.42   \\
3-0     &       -0.2    &       53.29   &       7.55    \\
3-1     &       -0.1    &       30.66   &       8.77    \\
3-2     &       -0.2    &       62.78   &       3.77    \\
4-0     &       -1.6    &       69.89   &       8.14    \\
4-1     &       -0.3    &       42.20   &       16.48   \\
4-2     &       -1.3    &       73.59   &       13.37   \\
4-3     &       -0.8    &       90.81   &       11.69   \\
5-0     &       -1.2    &       54.55   &       18.68   \\
5-1     &       -2.0    &       61.64   &       13.50   \\
5-2     &       -0.5    &       44.98   &       16.22   \\
5-3     &       -1.3    &       66.96   &       6.48    \\
5-4     &       -1.2    &       86.17   &       7.98    \\
\hline\hline
\end{tabular}
\end{center}
\end{table}

\begin{table}
\begin{center}
\caption{HF reduced-potential fitting parameters. The lower states $j'$, the optimized values of $C$,
the change in $C$ ($\Delta C$), and the slope of the derived linear
functions.
\label{table3}}
\vskip 0.5cm
\begin{tabular}{c c c c }
\hline \hline
$j'$   & C & $\Delta C$   & Slope \\
\hline
0       &       1.7     &       1.1     &       -0.94   \\
1       &       0.6     &       0.8     &       -0.72   \\
2       &       -0.2    &       0.6     &       -0.56   \\
3       &       -0.8    &       0.4     &       -0.44   \\
4       &       -1.2    &       0.2     &       -0.36   \\ \hline  
5       &       -1.4    &       0       &       -0.32   \\
6       &       -1.5    &       0       &       -0.30   \\
7       &       -1.6    &       0       &       -0.28   \\
8       &       -1.7    &       0       &       -0.26   \\
9       &       -1.8    &       0       &       -0.24   \\
10      &       -1.9    &       0       &       -0.22   \\
11      &       -2      &       0       &       -0.20   \\
12      &       -2.1    &       0       &       -0.18   \\
13      &       -2.2    &       0       &       -0.16   \\
14      &       -2.3    &       0       &       -0.14   \\
15      &       -2.4    &       0       &       -0.12   \\
16      &       -2.5    &       0       &       -0.10   \\
17      &       -2.6    &       0       &       -0.08   \\
18      &       -2.7    &       0       &       -0.06   \\
19      &       -2.8    &       0       &       -0.04   \\
20      &       -2.9    &       0       &       -0.02 \\
\hline\hline
\end{tabular}
\end{center}
\end{table}

\vskip 2.0cm

\vskip 2.0cm

\vskip 2.0cm

\vskip 2.0cm

\vskip 2.0cm

\end{document}